\documentclass[11pt,twoside]{book}
\usepackage{konkolyproc2}
\usepackage{longtable}
\usepackage{amsmath,amssymb}
\usepackage{graphicx}
\usepackage{lscape}
\usepackage{index}
\usepackage{natbib}
\usepackage{bigdelim}
\usepackage{multirow}

\usepackage{hyperref}
\usepackage{physics}
\usepackage{subfig}

\makeindex

\begin{document}

\pagestyle{myheadings}
\setcounter{equation}{0}\setcounter{figure}{0}\setcounter{footnote}{0}\setcounter{section}{0}\setcounter{table}{0}\setcounter{page}{1}
\markboth{Bellinger, Wysocki \& Kanbur}{Measuring amplitudes of variable stars}
\title{Measuring amplitudes of harmonics and combination frequencies in variable stars}
\author{Earl P. Bellinger$^{1,2}$, Daniel Wysocki$^3$ \& Shashi M. Kanbur$^4$}
\affil{$^1$Max-Planck-Institut f\"ur Sonnensystemforschung, G\"ottingen, Germany\\
$^2$Stellar Astrophysics Centre, Aarhus, Denmark\\
$^3$Rochester Institute of Technology, NY, USA\\
$^4$State University of New York at Oswego, NY, USA}

\begin{abstract}
Discoveries of RR Lyrae and Cepheid variable stars with multiple modes of pulsation have increased tremendously in recent years. The Fourier spectra of these stars can be quite complicated due to the large number of combination frequencies that can exist between their modes. As a result, light-curve fits to these stars often suffer from undesirable ringing effects that arise from noisy observations and poor phase coverage. These non-physical overfitting artifacts also occur when fitting the harmonics of single-mode stars as well. Here we present a new method for fitting light curves that is much more robust against these effects. We prove that the amplitude measurement problem is very difficult (NP-hard) and provide a heuristic algorithm for solving it quickly and accurately. 
\end{abstract}

\section{Introduction to the Fourier Decomposition}
The light curve of a Cepheid or RR Lyrae variable star with one or more modes of pulsation can be represented as a sum of periodic components:
\begin{equation}
    m(t ; \vec \omega, \mathbf A, \boldsymbol \Phi) = \sum_{k_1={-N}}^N \ldots \sum_{k_{|\vec \omega|}={-N}}^N A_{\vec k} \sin(t\qty[\vec k\cdot \vec \omega] + \Phi_{\vec k})
\end{equation}
where $t$ is the time of observation, $m$ is the magnitude, $\vec k$ is a vector of wavenumbers, $\vec \omega$ is a vector of angular frequencies, $\mathbf A$ is a multidimensional array of amplitudes, and $\boldsymbol \Phi$ is a multidimensional array of phases. This equation is known as the \emph{Fourier decomposition} and is especially useful for fitting light curves of stars that have been sampled irregularly in time.

The ordinary approach of measuring the amplitudes and phases in this equation begins by first separating each component into a sum of sines $\mathbf S$ and cosines $\mathbf C$ with
\begin{equation} \label{eq:varsep}
    A_{\vec k} \sin(t\qty[\vec k\cdot \vec \omega] + \Phi_{\vec k}) = S_{\vec k} \sin(t\qty[\vec k\cdot \vec \omega]) + C_{\vec k} \cos(t\qty[\vec k\cdot \vec \omega]).
\end{equation}
A matrix $\mathbf{X}$ is constructed containing columns for each sine and cosine term and a row for each individual observation. The amplitude of each component of the Fourier fit can then be measured with least squares linear regression, i.e.~ $\qty[\mathbf{S}\; \mathbf{C}] = (\textsf{\textbf{X}}^\text{T} \textsf{\textbf{X}})^{-1} \textsf{\textbf{X}}^\text{T} \vec{m}$, and finally we can obtain
\begin{equation}
    A_{\vec{k}} = \sqrt{C_{\vec{k}}^2 + S_{\vec{k}}^2} \quad \text{ and } \quad \tan(\Phi_{\vec{k}}) = -S_{\vec{k}} / C_{\vec{k}}.
\end{equation}
The only thing left to be determined is the order of fit $N$, that is, the number of components needed to describe the signal. This is commonly achieved by successive prewhitening or with the procedure known as Baart's criterion, which involves iteratively increasing $N$ until the auto-correlation of the residuals are below some threshold \citep{baart1982use, petersen1986studies}. 

\begin{figure}[!t]
    \centering
    \includegraphics[width=\textwidth]{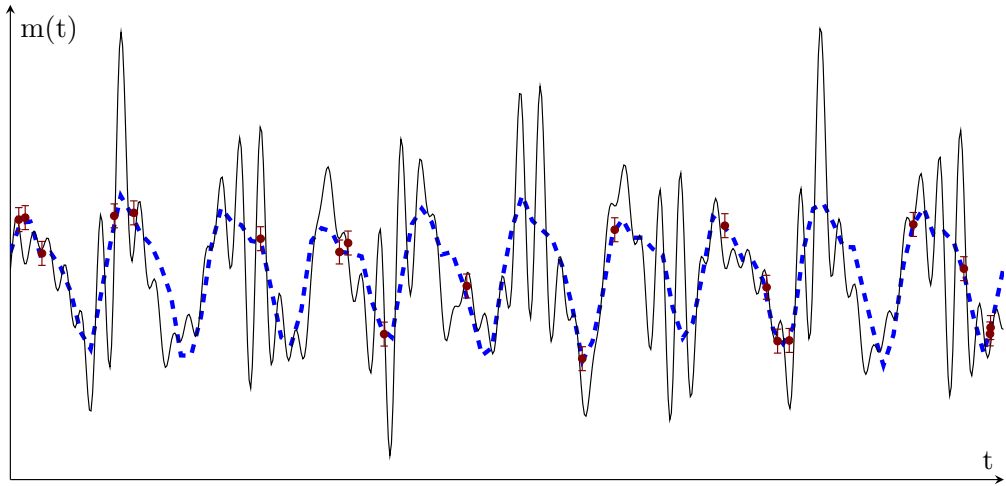}
    \caption{An example of a simulated multi-mode variable star being fit by the least squares Fourier decomposition. The blue dotted line is the true light curve of the star, the red points are the observations, and the solid line is the least squares fit. The fit is very poor and shows strong non-physical ringing effects.} 
    \label{fig:badfit} 
\end{figure}

Unfortunately, these procedures can result in very poor fits to observational data, especially when the data contain outliers or the time series has significant gaps in the phase coverage of the periods. An example of this can be seen in Fig.~\ref{fig:badfit}, in which a time series of observations for a simulated multi-mode variable star is fitted using the least squares Fourier decomposition. 

\section{Improving the Fourier Decomposition with Regularization}
We want to estimate the optimal parameters $\mathbf {\hat A}$ and $\boldsymbol{\hat \Phi}$ for a multi-mode oscillator $\hat m$ that is best supported by the observed data $\qty(\vec t, \vec m, \vec \epsilon)$, where $\vec \epsilon$ are the uncertainties on the observations. We also want to find the simplest model; that is, the one with the fewest components needed to describe everything we witnessed. This is just Occam's razor. And finally, we want to minimize the squared error between our model and the observations with the uncertainty on each observation taken explicitly into account. Putting this all together, we have
\begin{equation}
    \qty(\mathbf {\hat A}, \boldsymbol{\hat \Phi}) =
      \underset{\qty(\mathbf A, \boldsymbol \Phi)}{\arg\min} 
      \left(\left\|\mathbf{A} \right\|_0,
            \left\| \epsilon^{-1}\qty[\vec m - \hat m\qty(\vec t ; {\vec \omega}, \mathbf{A}, \boldsymbol{\Phi})]\right\|_2
      \right). 
\end{equation}
This optimization problem has several aspects that make it very difficult to solve. Not only does it have multiple objectives, but it is also a sparse ($\ell_0$-norm) minimization problem, which is NP-hard \citep{natarajan1995sparse} and therefore not able to be solved in practice. Hence, we are required to simplify the problem. 

\noindent If we relax the constraint on the amplitudes to $\ell_1$-norm minimization (that is, minimizing $\left\|\mathbf{A} \right\|_1$), which encourages sparsity rather than requiring it, then we can make use of the method of Lagrangian multipliers to scalarize the objectives. We can combine the objective functions with a \emph{regularization parameter} that can be chosen either via cross-validation or with an information criterion such as Akaike (\citeyear{akaike1974new}) or Bayes \citep{schwarz1978estimating}. This is equivalent to putting Laplacian priors on all of the amplitudes. This simplified problem is known in regression analysis as the Least Absolute Shrinkage and Selection Operator, or \emph{LASSO}, and we can solve it using quadratic programming \citep{tibshirani1996regression}, coordinate descent \citep{fu1998penalized}, or least-angle regression \citep{efron2004least}. 

\begin{figure}[!t]
    \centering
    \includegraphics[width=\textwidth]{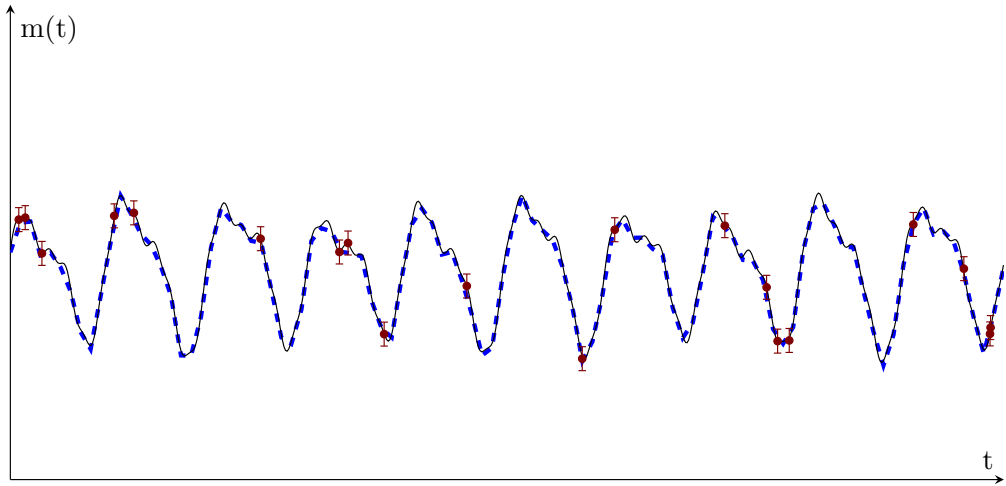}
    \caption{The same data in Fig.~\ref{fig:badfit} being fit with the LASSO method.}
    \label{fig:goodfit} 
\end{figure}

\begin{figure}[!ht]
    \centering
    \includegraphics[width=\textwidth,keepaspectratio]{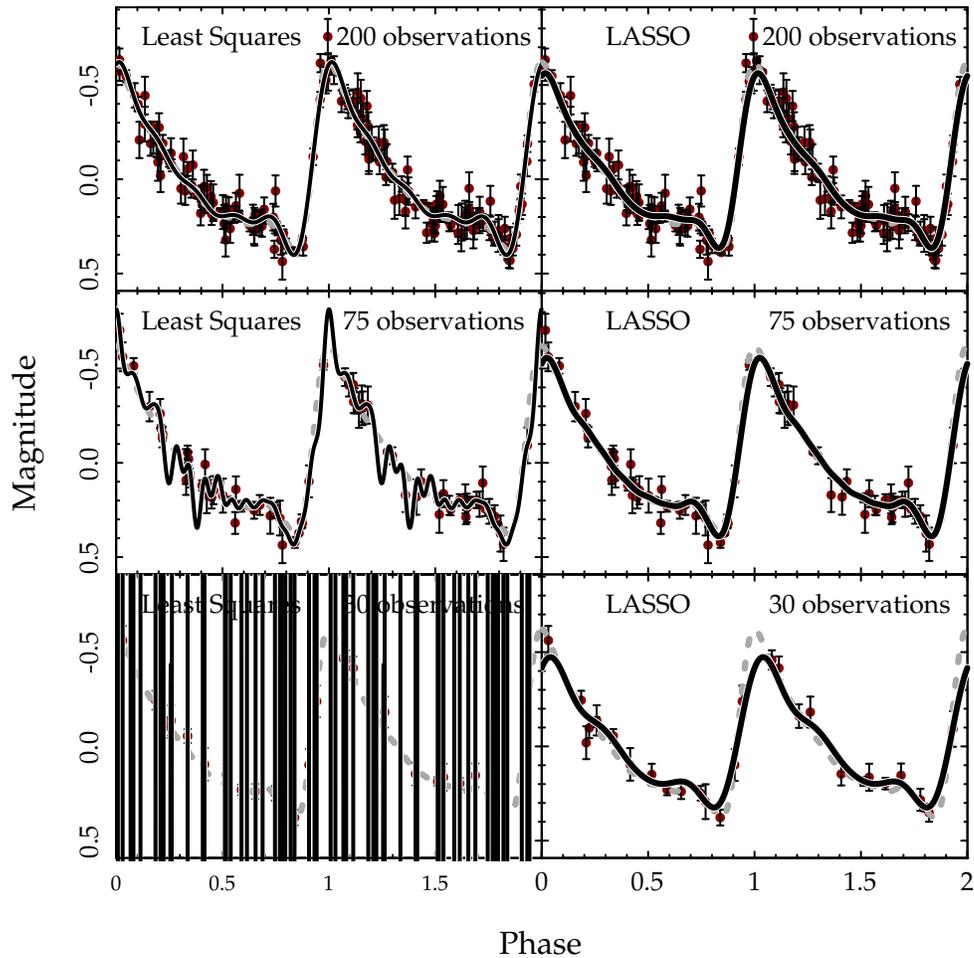}
    \caption{Sensitivity analysis of a simulated RR Lyrae light curve (dashed gray line). When the number of observations (red points) is large (top), both least squares (solid black line, left) and LASSO (solid black line, right) fits perform well. When the number of observations is small (bottom), however, the least squares fit fails catastrophically and only LASSO still works as desired.} 
    \label{fig:sensitivity} 
\end{figure}

In Fig.~\ref{fig:goodfit}, we return to the simulated multi-mode variable star and fit it with our LASSO method. It can be seen that the ringing effects have been eliminated. In Fig.~\ref{fig:sensitivity}, we apply this method to a classical RR Lyrae light curve with one period of pulsation. Here it can be seen that the LASSO vastly outperforms the least squares method, especially when there are few data points. 

We are developing a free, open source, and easy-to-use code for fitting light curves with the LASSO method. In addition, we are preparing a catalog of LASSO light-curve fits for stars that have been observed by OGLE-III. These will both be released in a future publication. The source codes for producing the figures in this manuscript are available electronically at \\ \url{https://github.com/earlbellinger/multiperiod} \citep{bellinger2015lasso}.

\section*{Acknowledgements}
EPB thanks ERC grant agreement no.~338251 (StellarAges), the NPSC and NIST for fellowship support, and the Max Planck Society and IMPRS for travel support. The authors thank the IUSSTF for their support of the Joint Center for the Analysis of Variable Star Data that funded collaborative visits. DW thanks SUNY Oswego for support to visit the University of Delhi. 

\end{document}